\documentstyle[epsfig,12pt]{article}

\begin{document}

\vspace{.5in}
\setlength{\parindent}{1in}
\begin{center}
{\Large The Electromigration Force in Metallic Bulk}

\vspace{.2in}
{\large A. Lodder and J.P. Dekker}

\vspace{.2in}
{\small
{\it Faculteit Natuurkunde en Sterrenkunde, Vrije Universiteit,\\
De Boelelaan 1081, 1081 HV Amsterdam, The Netherlands\\}}
\end{center}
\vspace{.3in}

\setlength{\parindent}{8.5mm}


{\small
{\bf Abstract.} The voltage induced driving force on a migrating atom in a
metallic system is discussed
in the perspective of the Hellmann-Feynman force concept, local screening
concepts and the linear-response approach.
Since the force operator is well defined in quantum mechanics it
appears to be only confusing to refer to the Hellmann-Feynman theorem
in the context of electromigration. Local screening concepts
are shown to be mainly of historical value. The physics
involved is completely represented in
{\it ab initio} local density treatments
of dilute alloys and the implementation does not require
additional precautions about screening, being typical for jellium treatments.
The linear-response approach is shown to be a reliable guide
in deciding about the two contributions to the driving force, the
direct force and the wind force.

Results are given for the wind valence for electromigration
in a number of FCC and BCC metals, calculated using an {\it ab initio}
KKR-Green's function description of a dilute alloy.
}

\begin{center}
\section*{\large{\bf INTRODUCTION}}
\label{sec:introduction}
\end{center}

Electromigration in a metal, being the drift of atoms under the influence of
an electric field, is interesting from different points of view
\cite{ho}.
From the theoretical side a question arose about the screening of the charge of
the migrating particle.
This question was the subject of a long-standing controversy, which
has been clarified  to a large extent only recently \cite{clarification,EL94}.
On the technological side the problem of the failure of integrated circuits, 
the so-called aluminum catastrophe, has attracted attention up to now. 
This problem has mainly been approached empirically, which has led to the
observation that the addition of copper impurities slows down the development
of voids.
For a long time this electromigration induced failure
was attributed to the high diffusion rate along 
grain boundaries. Recently, ongoing miniaturization has led to
interconnects with a bamboo structure \cite{lloyd}, in which interface and
bulk diffusion are becoming more and more important \cite{itsk}.

In any attempt to describe the process of electromigration one uses
the conventional decomposition of the driving force {\bf F}
in an electrostatic direct force, and a wind force attributed to
momentum transfer of the current carrying electrons to the
migrating atom, 
\begin{equation}
\label{eq:Feff}
{\bf F}={\bf F}_{\rm direct}+{\bf F}_{\rm wind}=
(Z_{\rm direct}+Z_{\rm wind})e{\bf E}=Z^{*}e{\bf E}.
\end{equation}
The forces are proportional 
to corresponding valences and the applied electric field {\bf E}.
The effective valence $Z^{*}$ is the measurable quantity.
A basic task of the theory of electromigration is to provide
with a microscopically correct expression for the
driving force. 

In a history of over thirty years this task
has not been achieved completely yet. This is certainly due to the
complexity of the phenomenon of electromigration. Although
the force is a local quantity, the presence of a current density
requires the treatment of a macroscopically large system.
Semiclasssical treatments, along with fully quantum-mechanical
ones, were supplemented by momentum-balance considerations,
the latter being obscured by a treacherous interpretation
dependence. Also the force according to the Hellmann-Feynman
theorem was considered \cite{sorhell,hellmann,feynman}. For a review of the various
approaches, including a clearly structured account of a new
linear-response result for the driving force, see Ref. \cite{EL94}.

Still, we think that due to new developments, both
on the fundamental level of deriving a reliable expression
for the driving force and on the practical level of
doing actual calculations of the wind valence
in real metals,
things have become much clearer than fifteen
years ago. We will illustrate that first by discussing
the Hellmann-Feynman theorem in relation to
the driving force. After that the merit of local concepts
such as the residual-resistivity dipole will be evaluated,
using linear-response theory and results from {\it ab initio} calculations.
Finally, results will be presented for the wind valence
in FCC and BCC metals.

\begin{center}
\section*{\large{\bf HELLMANN-FEYNMAN THEOREM AND DRIVING FORCE}}
\label{sec:hellmann}
\end{center}

An appealing expression for the driving force on an atom at position ${\bf R}_{p}$,
\begin{equation}
\label{eq:Fdens}
{\bf F}_{p} = Z_{\rm direct} e{\bf E} - 
\int \delta n({\bf r}) \nabla_{{\bf R}_{p}}v_{p}d^{3}r,
\end{equation}
is found in many papers \cite{sorhov,sormes}. Since $\delta n({\bf r})$ is the
deviation of the local electron density from the equilibrium density, 
being linear in the electric field,
the second term is viewed as an obvious form for
the wind force, $- {\nabla}_{{\bf R}_{p}}v_{p}$ being the operator for
the force on the atom.
This equation was already described by Bosvieux and
Friedel \cite{friedel}, but in their formalism $Z_{\rm direct}$ turned out to
be zero for an interstitial atom. The latter fact gave rise to the
controversy mentioned above, to which we will return later on.
Eq. (\ref {eq:Fdens}) is also reminiscent of
the force expression following from Feynman's theorem, dating
back to the early days of quantum chemistry calculations \cite{feynman}.
Here we want to add to the
discussion started by Sorbello and Dasgupta \cite{sorhell} about the
relationship between the two.

Feynman derived the equality
\begin{equation}  
\label{eq:Feynm1}
\nabla_{{\lambda}} {\cal E} ({\lambda}) = 
<\psi_{{\lambda}}|\nabla_{{\lambda}}H({\lambda})|\psi_{{\lambda}}>,
\end{equation}
in order to facilitate quantum-chemistry calculations of the
equilibrium configuration of atoms in a molecule. One was used
to calculate the ground state energy ${\cal E} ({\lambda}) =
<\psi_{{\lambda}}|H({\lambda})|\psi_{{\lambda}}>$ as a function of a
parameter ${\lambda}$, for example the ${\it x}$ component of the
position of an atom. By varying the value of ${\lambda}$ and repeating
the calculation for the ${\it y}$ and ${\it z}$ components,
one determined the direction of evolution toward
an equilibrium configuration. In this perspective
Eq. (\ref {eq:Feynm1}) is very helpful
in calculating the force at a position corresponding to a value
of ${\lambda}$, because one can restrict oneself to the calculation
of one matrix element for that value of ${\lambda}$, the matrix
element at the right-hand side of this equation.

Now consider a system of a free molecule under the influence of a
uniform electric field. In addition to the potentials $V_{j}$ of
the nuclei at positions ${\bf R}_{j}$, the Hamiltonian contains
a term $- \sum_{j} Z_{j} e {\bf R}_{j}.{\bf E}$ and a similar
term for the electrons interacting with the field. Applying
Eq. (\ref {eq:Feynm1}) for ${\lambda} = {\bf R}_{p}$, one finds
for the so-called Hellmann-Feynman force
\begin{equation}   
\label{eq:Feynm2} 
{\bf F}_{p}^{{\rm HF}} = - \nabla_{{\bf R}_{p}} {\cal E} ({{\bf R}_{p}}) = 
 Z_{p} e{\bf E} -
<\psi_{{\bf R}_{p}}|\nabla_{{\bf R}_{p}}V_{p}|\psi_{{\bf R}_{p}}>
+ \sum_{j({\not=}p)} {\bf F}_{pj}.
\end{equation}
The third term corresponds to the force on the nucleus at
the position ${\bf R}_{p}$ due to repulsion by the other nuclei. Since
$V_{p} =  \sum_{i} v^{0}_{p}({\bf r}_{i} - {\bf R}_{p})$ 
is a sum over all electrons, at positions
${\bf r}_{i}$, of the nuclear potential $v^{0}_{p}$, the many-body
matrix element in Eq. (\ref {eq:Feynm2}) can be written as a
one-body integral over the electron density, by which this equation
becomes \cite{sorhell}
\begin{equation}
\label{eq:Feynm3}
{\bf F}_{p}^{{\rm HF}} =
 Z_{p} e{\bf E} -\int n({\bf r},\lbrace{\bf R}_{j}\rbrace, {\bf E}) 
\nabla_{{\bf R}_{p}}v^{0}_{p}d^{3}r + \sum_{j({\not=}p)} {\bf F}_{pj}.
\end{equation}
For a molecule with the nuclei at equilibrium positions, the second term
with ${\bf E} = 0$ cancels the third one. Then the force 
$\delta {\bf F}_{p}^{{\rm HF}}$, purely induced by the applied fied, can be written as
\begin{equation}
\label{eq:Feynm4}
\delta {\bf F}_{p}^{{\rm HF}} = 
 Z_{p} e{\bf E} -\int \delta n({\bf r},\lbrace{\bf R}_{j}\rbrace, {\bf E})
\nabla_{{\bf R}_{p}}v^{0}_{p}d^{3}r,
\end{equation}
which is proportional to the field ${\bf E}$ and is most similar to
Eq. (\ref {eq:Fdens}). Sorbello et al. \cite{sorhell} evaluated
Eq. (\ref {eq:Feynm4}) for an isolated one-electron atom and showed
explicitly that the force cancels, as it should. The force (\ref {eq:Feynm4})
remains finite for a nucleus in a molecule. Only the
total force, the sum $\sum_{p} 
\delta {\bf F}_{p}^{{\rm HF}}$, 
cancels. 

The Hellmann-Feynman theorem was derived for a finite
system such as a molecule. Although electromigration occurs in a
current carrying solid, due to the similarity of Eqs.
(\ref{eq:Fdens}) and (\ref{eq:Feynm4}) one might wonder whether still
there is a relation between the two. In order to understand why
the final answer is negative, one has
to analyze the ingredients of the
theorem (\ref{eq:Feynm1}). First, the system is bound to be
finite because the total energy ${\cal E}(\lambda)$ is the
starting point. Secondly, and we think more importantly,
a single ground state ${\psi}_{\lambda}$ is supposed to
be available or calculable. 

Since most of the theories for
solids use a finite volume and periodic boundary conditions,
the first limitation is not serious. That is why
one still refers to it \cite{zeller} in doing
electrostatics calculations on lattice distortion around an
impurity. In taking the thermodynamic limit, {\it i.e.}
going to infinite volume and keeping the density constant,
the volume factors drop out and the summations turn into
integrals, which, ironically, in present days are evaluated
numerically as summations. In electrostatics calculations
one effectively operates at absolute zero temperature,
the Fermi surface is sharp and one can work with the
concept of a ground state. In electromigration, however,
and in dealing with any transport property, this is
obviously not the case. One has to work at a finite
temperature. The system cannot be described with
a single wave function ${\psi}_{{\bf R}_{p}}$. A
macrocanonical ensemble of states has to be used. One has
to turn to the means developed in non-equilibrium
quantum statistical mechanics. To that end a Kubo-formula like
linear-response expression for the driving force was
proposed by Kumar and Sorbello \cite{kumar}.

It seems legitimate to ask what nowadays is still the
message of the Hellmann-Feynman theorem (\ref {eq:Feynm1}),
particularly in calculating local forces in systems. For
that purpose
the equality (\ref {eq:Feynm1}) seems to be a trivial one.
We consider it as a possible answer, that in
the early days of quantum mechanics
a force operator was more or less suspect, reminding too much of
newtonian classical mechanics. A Hamiltonian and the energy were considered
as reliable objects. Nowadays we know that the force
operator is a reliable object as well, and Eq. (\ref {eq:Feynm1})
just confirms that fact.

Taking things together, referring to Eq. (\ref {eq:Feynm1})
in electrostatics calculations is not necessary, because
one actually calculates the expectation value of the
force operator. Authors still continue to do so, but it
merely serves as a label that physicists recognize. Referring
to the Hellmann-Feynman theorem in electromigration theory
is even confusing. One has to concentrate on the evaluation of the
force operator in a statistical ensemble, while the
Hellmann-Feynman theorem deals with a single ground
state only, and does not give a clue how to go beyond that.

We conclude this section by a discussion of the
linear-response expression for the driving force, which
can be derived from the quantum-statistical expectation
value of the force operator, directly coming from the commutator
of the momentum operator ${\bf P}_{p}$ with the total Hamiltonian,
\begin{equation}
\label{eq:lr1}
{\bf F}_{p}= {\rm Tr} \lbrace \rho \frac {d{\bf P}_{p}}{dt} \rbrace =
Z_{p} e {\bf E} - {\rm Tr}\lbrace \rho \nabla_{{\bf R}_{p}}V_{p}\rbrace.
\end{equation}
The  many-body linear-response form
can be simplified to the form \cite{lodder1}
\begin{equation}
\label{eq:lr2}
{\bf F}_{p} = Z_{p} e{\bf E} - {\rm i} e \lim_{a \rightarrow 0} {\int}^{\infty}_{0}
 dt {e}^{-at} {\rm tr} \lbrace n(h) [{\bf f}_{p}(t), {\bf r}\cdot{\bf E}]\rbrace,
\end{equation}
for any system with an unperturbed Hamiltonian $H$ that
can be written as a sum of single-electron Hamiltonians
\begin{equation}
\label{eq:hsum}
H = \sum_{i} h^{i}.
\end{equation}

The latter property holds for the electron-impurity model
system described in the local density approximation, its
most simple realization being the impurity-in-a-jellium
model. In such a model system electron-phonon interaction
is not accounted for, but it is good to realize, that
almost all evaluations of Eq. (\ref {eq:lr2}) or its
many-body form pertain merely to the electron-impurity
model system.
In Eq. (\ref {eq:lr2}) the time dependence of the force
operator ${\bf f}_{p} = - \nabla_{{\bf R}_{p}} v_{p}$ is given by
\begin{equation}
\label{eq:fpt}
{\bf f}_{p} (t) = e^{{\rm i}ht} {\bf f}_{p} e^{-{\rm i}ht},
\end{equation}
while $n(h)$ is the Fermi-Dirac distribution in operator form
\begin{equation}
\label{eq:nh}
n(h) = \frac{1}{e^{\beta(h - \epsilon_{F}}) + 1},
\end{equation}
$\epsilon_{F}$ being the Fermi energy and ${\beta}^{-1} = k_{B}T$.
Comparing Eq. (\ref {eq:lr2}) with Eq. (\ref {eq:Fdens}),
one is inclined to identify the second term in Eq.
(\ref {eq:lr2}) as the wind force. In evaluating the
trace operator in the linear-response expression,
Sham \cite{sham} already found that this identification
holds only partially. In addition to the real
wind force he found a screening contribution,
to be combined with the unscreened first term
in Eq. (\ref {eq:lr2}). 

Sham evaluated a many-body expression, but
his result can be read quite simply from Eq. (\ref{eq:lr2}),
which has been proven to be equivalent to Sham's expression \cite{lodder1}.
The real wind force follows from the second term in this equation
after the replacement of the one-electron system Hamiltonian,
\begin{equation}
\label{eq:hh0v}
h = h_{0} + \sum_{j} v_{j},
\end{equation}
in the statistical factor $n(h)$, by its unperturbed
part $h_{0}$, $v_{j}$ being the electron-impurity
potential for the impurity at the position ${\bf R}_{j}$.
Sham proved this to lowest order in
the electron-impurity potential, and it appears to hold
to all orders, which are
generated by the time dependent factors around the
force operator \cite{lodder2}. The screening contribution comes
from what is left, namely a term with $n(h) - n(h_{0})$
instead of $n(h)$. To lowest order in the electron-impurity
potential Sham found a negligible screening contribution.
Recently \cite{lodder1} his result has been confirmed, but in addition
it was shown to be possible to evaluate the expression
to all orders. That analysis leads to a complete cancellation of the
direct force.  Some objections were raised \cite{landauer0,sorbello0},
but it was not difficult to unravel them \cite{lodlan,lodsor}. Although the
proof is valid only for the electron-impurity system, it
settles an important point in the controversy. Sham's result has been
used again and again in arguing in favour of an unscreened 
direct force \cite{sormes,adver}. The argument was, that a lowest
order result, being quadratic in the potential, and therefore
quadratic in $Z_{p}$, logically excludes a complete cancellation
of the linear unscreened force. 

Unfortunately, experiments on hydrogen in V, Nb and Ta, set up
to decide in this matter, turned out to be not conclusive
\cite{AdVerGr}. Although complete screening could be excluded, for
hydrogen in Nb the fitting procedure resulted in
the rather small direct valence value of 0.44.
An attempt to acccount for electron-phonon interaction
points in the direction of a reduced screening on
increasing the temperature \cite{clarification,lodder3}.
Since this attempt is not an exact result, the question
about the screening of the direct force cannot
be considered as completely settled yet.

Local screening-field approaches will be discussed
in the following section.

\begin{center}
\section*{\large{\bf LOCAL CONCEPTS AND DRIVING FORCE}}
\label{local}
\end{center}

Parallel to the formulation of the driving force according
to a Kubo formula approach \cite{kumar,lodder1,sham}, a local
screening field approach has been applied, inspired by
Landauer \cite{landauer1,landauer2,landauer3}. Illustrative
examples of recent work are the treatment of electromigration
in mesoscopic systems by Sorbello \cite{sormes} and a direct
force calculation by Ishida \cite{ishida}. Landauer's concepts
leading to the wind and direct forces are the residual
resistivity dipole \cite{landauer2,sorres} and
carrier density modulations \cite{landauer3,ishida} respectively.
Although it is accepted that these concepts cover an aspect
of the physical reality of electronic behaviour in the neighbourhood
of an impurity in a metal \cite{sorres}, it is admitted that
they neither have led to new answers for the residual
or impurity resistivity \cite{sorres,sorchu}, nor have
contributed to the resolution of the controversy about
the driving force in electromigration. In the present section
we want to comment on recent work \cite{sormes,ishida,kaxiras},
keeping in mind the result of complete screening of the
direct force on an impurity in a jellium, obtained using
a Kubo-formula approach \cite{lodder1}.

In dealing with electromigration in mesoscopic systems,
Sorbello \cite{sormes} finds an expression for the wind force
equivalent to the second term in Eq. (\ref{eq:Fdens}) and
completely in agreement
 with the original Bosvieux-Friedel \cite{friedel}
expression and later generalizations of it used in metallic
systems \cite{sorhov}. His expression  for the direct force
contains two terms, an unscreened term and a term 
accounting for screening, which composition is in agreement
with the Kubo-formula analyses discussed above
\cite{lodder1,sham}.
His method is based on treating local transport fields at
and around an impurity immersed in a ballistic
mesoscopic environment, being essentially a
jellium environment. Electron-phonon interaction
is negligible, because in his model systems the
inelastic electron mean-free-path exceeds the
system size. Screening is dealt with according to well established
jellium rules \cite{ziman}.

For two of the three model systems considered he finds
a vanishing direct force.
In evaluating the direct force in his third 
model system, an impurity in the vicinity of a
point contact, he does not find complete screening.
It is most interesting to observe, that in obtaining
this latter result an argument was used derived from
a consensus interpretation of
incomplete linear-response treatments \cite{sham,rimby,sor85} ,
the presently available complete treatment not being
published yet at that time \cite{lodder1,lodder4}. 
That may explain the struggle
of the author with what he calls a paradox, and his
inclination to distrust the Kubo approach.

Ishida \cite{ishida} studies the simple system of one
impurity in a jellium, and finds a small screening only.
This contradicts the exact Kubo-formula result \cite{clarification,lodder2}
and is not in agreement with Sorbello's findings either \cite{sormes}.
An objection as that in the Kubo-formula approach the screening is
not accounted for self-consistently, is not valid. By
starting from the complete Hamiltonian and evaluating the
exact formal expression, {\it all} dynamics in the system
is accounted for. The cause of Ishida's different result
was uncovered in a discussion with the author \cite{ishidaL}.
Ishida uses an {\it ac} field in order to avoid infinities in
his wind force. However, the infinitisimally small {\it a} as
it occurs in Eq. (\ref{eq:lr2}), which comes from the
adiabatic switching-on of the electric field and plays the role
of the inverse relaxation time if no dissipative scattering
is present, was not treated with sufficient care. Although Ishida
is speaking about a small frequency $\omega$, all his results
are in the large frquency domain, because $\omega/a$ is always
much larger than unity, irrespective how small $\omega$ is.
In the $\omega \rightarrow 0$ limit, so in the true {\it dc}
limit, his formalism also leads to complete screening.

Kandel and Kaxiras \cite{kaxiras} calculated the direct force
on a Si atom on top of a Si surface. In this semiconductor problem
the wind force turns out to be negligible. Through a self-consistent
calculation they find a very small direct valence of 0.05, but
it is significantly different from zero. In a semiconductor
environment, with dangling bonds, a complete screening is
not expected {\it a priori}, but the actually calculated
direct valence is very small indeed.

Taking everything together, more and more results are pointing into
the direction of a screened direct force for an impurity
in an electron gas. Still, it remains true that this point has
not completely been settled yet, even apart from a possible reduced
screening due to electron-phonon interaction \cite{clarification,lodder3}.

Now let us turn to the wind force, which is equally present in
all treatments. In discussing the meaning of the residual
resistivity dipole (RRD), it is useful to introduce first
the well-established expression used in practical calculations
for metallic systems \cite{sorhov,sormes,1984,ek}
\begin{equation}
\label{eq:WFexpr}
{\bf F}_{{\rm wind}}=\sum_{{\bf k}} \delta f({\bf k})
<\Psi_{{\bf k}}|-\nabla_{{\bf R}_{p}}v_{p}|\Psi_{{\bf k}}>.
\end{equation}
The deviation $\delta f({\bf k})$ from the Fermi-Dirac distribution,
here denoted by $f_{0}(\epsilon_{\bf k})$ instead of by $n(\epsilon_{\bf k})$
(see Eq. \ref {eq:nh}),
\begin{equation}
\label{eq:deltaf}
\delta f({\bf k})=e \tau {\bf E} \cdot {\bf v}_{\bf k}
\frac{d}{d\epsilon_{{\bf k}}}f_{0}(\epsilon_{{\bf k}})
\end{equation}
is a result of the applied electric field ${\bf E}$. The sum over
the states, labeled by {\bf k}, can be written as $\delta n({\bf r})$,
after which one recognizes the second term in the basic expression
(\ref {eq:Fdens}).
So, the wind force is proportional to the mean free transport time $\tau$
of electrons and therefore inversely proportional to the resistivity of the
sample.
The electronic structure is represented by the wave function $\Psi_{{\bf k}}$
and the scattering is taken care of by the potential $v_{p}$ of the migrating atom.

In applications to dilute alloys \cite{ek} the wave function $\Psi_{{\bf k}}$
is calculated using multiple-scattering theory. The electronic structure of
the unperturbed metallic host is calculated by the
Korringa-Kohn-Rostoker (KKR) method \cite{k,kr}. Since $\Psi_{{\bf k}}$
is the exact local wave function the electronic structure of the
electromigration defect has to be calculated, consisting of the migrating atom,
its near surroundings, possibly being affected by charge transfer
and lattice distortion, and, in the case of substitutional electromigration,
a neighbouring vacancy. For that purpose an advanced application of
multiple-scattering theory is used, called KKR Green's function theory
\cite{1984,ek,VEDL95}.
It may be useful to remark, that the wind force given 
by Eq. (\ref{eq:WFexpr})
depends on the position ${\bf R}_{p}$ of the migrating
atom along its jump path. 
This holds irrespective whether one considers migration of
an interstitial or a substitutional atom, the latter requiring a
neighbouring vacancy. The actually measured force ${\bf F}$ is calculated
from the work done by the position dependent force ${\bf F}_{p}$ along
the path, divided by the path length, and averaged over the possible
orientations of the path with respect to the ${\bf E}$ direction.
In the following section {\it ab initio} results will we given for $Z_{{\rm wind}}$
calculated that way, for a number of FCC and BCC metals.

\begin{figure}[t]
     \epsfig{figure=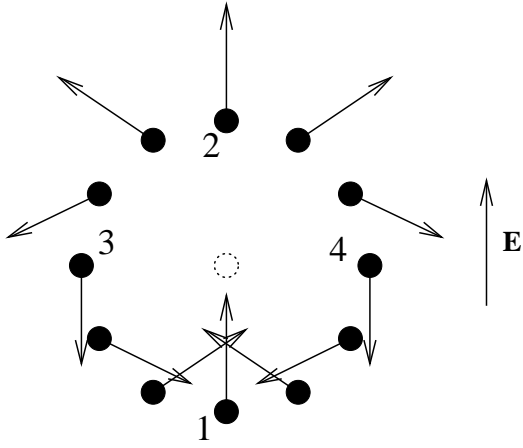,height=6.0cm,width=7.0cm}
     \caption[]{
          Wind force on a V atom next to a vacancy, indicated by a dotted
          circle, for different orientations of the 
          migration path with respect to the electric field {\bf E}.
          For position 1 the field lies along the migration path. For
          positions 3 and 4 it points perpendicular to it.}
     \label{fig:Als}
\end{figure}

Here we want to give a special result related to the RRD. In Fig.
\ref{fig:Als} the wind force is shown as it acts on vanadium atoms around
a vacancy in V.
Position 1 is the initial position 
in an actual migration jump. One sees that, going from position 1,
via the perpendicular positions 3 and 4, to position 2, so looking
around half a circle, the direction of the force turns over $360^{{\circ}}$.
Interpreting the force direction as showing the local current direction,
the vacancy apparently induces a back flow. Such an effect is
covered by the concept of the RRD. This can be seen as follows.

In discussing the RRD always an impurity in a free-electron medium,
or, equivalently, a jellium is considered, while the dissipative
property of the medium is represented by a transport relaxation
time \cite{sorres,sorchu}.
Following Sorbello \cite{sorres} in a slightly generalized form,
the wave function $\Psi_{{\bf k}}$ can be written as
\begin{equation}
\label{eq:psik}
\Psi_{{\bf k}}({\bf r}) = \frac{4 \pi}{\surd \Omega} \sum_{\ell m}{\rm i}^{{\ell}}
Y_{\ell m}(\hat k) R_{\ell}(r)Y_{\ell m}(\hat r),
\end{equation}
${\Omega}$ being the volume of the system, and the radial wave function
having the following asymptotic behaviour
\begin{equation} 
\label{eq:Rlr}
R_{\ell}(r) = j_{\ell}(kr) + {\rm i} \sin\delta_{\ell}e^{{\rm i}\delta_{\ell}}
h^{+}_{\ell}(kr).
\end{equation} 
Here  $j_{\ell}$ and $h^{+}_{\ell}$ are the spherical Bessel and Hankel
functions respectively, and $\delta_{\ell}$ are 
the phase shifts of the scattering potential.
It is clear that the Bessel function part of $\Psi_{{\bf k}}({\bf r})$
sums precisely up to the normalized plane wave $e^{{\rm i}{\bf k}.{\bf r}}/
{\surd \Omega}$. Using that the derivative of the Fermi-Dirac distribution
$f_{0}(\epsilon_{\bf k})$ with respect to the energy $\epsilon_{\bf k}$
effectively is equal to $- \delta (\epsilon_{\bf k} - \epsilon_{F})$,
and restricting oneself to {\it s} scatterers, one finds, that
$\delta n({\bf r})$, according the Eqs. (\ref{eq:Fdens}) and
(\ref{eq:WFexpr}) given by
$\sum_{{\bf k}} \delta f({{\bf k}})|\Psi_{{\bf k}}({{\bf r}})|^{2}$, becomes equal to
\begin{equation}
\label{eq:dnr}
\delta n({\bf r}) = 
A {\bf E} \cdot{\hat r} \sin\delta j_{1}(kr)[\cos\delta j_{0}(kr)
 - \sin\delta n_{0}(kr)]
\equiv A {\bf E} \cdot{\bf r} g(r),
\end{equation}
in which {\it A} is a positive constant equal to $2e\tau k_{F}^{2}/\pi^{2}$.
Only the ${\ell} = 0$ phase shift, 
denoted by ${\delta}$, is left. This expression clearly has a dipolar
form. Sorbello \cite{sorres} even distinguishes a Bosvieux-Friedel dipole, given
by the first term,  and a RRD dipole proportional to $\sin^{2}\delta$.

Now the wind force can be calculated for a configuration as in
Fig. \ref{fig:Als}. Modeling the vacancy by a strong {\it s} scatterer,
and a surrounding atom by a weak potential, centered at {\bf R},
the following integral
\begin{equation}
\label{eq:FvR}
{\bf F}_{{\rm wind}} = - A \int {\bf E} \cdot{\bf r} g(r) \nabla_{{\bf R}} 
v({\bf r} - {\bf R})
d^{3}r = - A\nabla_{{\bf R}}\int {\bf E} \cdot{\bf r} g(r) 
v({\bf r} - {\bf R}) d^{3}r
\end{equation}
has to be evaluated. A most simple, but
illustrative result is obtained for an attractive delta function potential
$v({\bf r} - {\bf R}) = -  C \delta({\bf r} - {\bf R})$ with a small positive
constant {\it C}. One finds
\begin{equation} 
\label{eq:delta}
{\bf F}_{{\rm wind}} = AC [{\bf E} g(R) + ({\bf E} \cdot{\hat R}) {\bf R} g'(R)].
\end{equation}
This force has a behaviour which is similar to what is depicted
in Fig. \ref{fig:Als}. The first term points along the field and is
proportional to $g(R)$. The second term, being proportional to
$Rg'(R)$, points along ${\bf R}$, but its direction changes sign,
and so turns over 180$^{\circ}$,  when ${\bf R}$ is perpendicular to ${\bf E}$.
So its direction turns over 360$^{\circ}$ while ${\bf R}$ turns over 180$^{\circ}$.
The precise behaviour depends on the ratio
$R \frac{d}{dR}\ln g(R)$. A typical value of this ratio is $-2$,
by which the second term dominates at the positions 1 and 2.
Due to its negative value the forces according to Eq. (\ref{eq:delta})
have a direction opposite to the forces in
Fig. \ref{fig:Als}. This is consistent with a jellium treatment, which
can give a negative $Z_{\rm wind}$ only.
\begin{table}[t]
\caption{Calculated $Z_{\rm wind}$  and some measured Z* values at 0.9 $T_{\rm m}$.}
\label{data}
\begin{tabular}{ccc|ccc}
\hline
\hline
\multicolumn{2}{r}{FCC} &&  \multicolumn{3}{c}{BCC}\\
system  &  $Z_{\rm wind}$ &  $Z^{*}$ &system  &  $Z_{\rm wind}$ &  $Z^{*}$ \\
\hline
Al      &  -3.11          &  -3.4    & V      &    0.99         &          \\
Al(Cu)  &  -5.29          &  -6.8    & Nb     &    0.76         &  1.3     \\
Al(Pd)  &  -8.71          &          & Ta     &    0.35         &          \\
Al(Si)  & -24.26          &          & Nb(Y)  &    0.95         &          \\
Cu      &  -3.87          &  -5      & Nb(Zr) &    0.93         &          \\
Ag      &  -3.51          &  -8      & Nb(Mo) &    0.85         &          \\
Ag(Pd)  &  -5.00          &          & Nb(Tc) &   -0.27         &          \\
Ag(Sn)  & -24.10          &          & Li     &   -4.18         &  -1.2    \\
Ag(Sb)  & -44.82          & -115     & Na     &   -0.28         &  -3.3    \\
Au      &  -3.77          &  -8.5    & K      &   -0.36         &          \\
Y       &   0.45          &          & Rb     &   -0.53         &          \\
Zr      &  -0.42          &  0.3     & Cs     &   -5.47         &          \\
Tc      &   0.44          &          & Ca     &    5.42         &          \\
Ru      &   0.07          &          & Sr     &    0.49         &          \\
Rh      &  -0.77          &          & Ba     &    0.93         &          \\
Pd      &  -1.33          &          & Ti     &    1.05         &          \\
        &                 &          & Zr     &    0.12         &  0.3     \\
        &                 &          & Cr     &    0.26         &          \\
        &                 &          & Mo     &    0.42         &          \\
        &                 &          & W      &    0.02         &          \\
\hline
\hline
\end{tabular}
\end{table}

Although existing evaluations of the RRD are given for an impurity
in a jellium only, Fig. \ref{fig:Als} shows that a RRD is
also present in a real solid. Nevertheless, in doing actual
calculations of the wind force the concept of a RRD does not enter
at all. One might wonder what is the reason. As it has
already been indicated earlier \cite{lodder2,sorres}, the RRD is accounted
for automatically in a complete multiple-scattering calculation. The
concept does not enter the formulation. It can be considered as
a manifestation of Landauer's physical intuition that he pointed at
such local effects, higher than to second order in the potential,
before one devised the means to make them explicit in a complete
scattering theory \cite{landauer1}. Further, the concept is helpful
in resolving a paradox in the calculation of the impurity resistivity
using statistical mechanics. In the latter treatment only
eleastic scattering enters, and dissipation cannot be understood.
The RRD makes it clear that a local field is building up.
But in practice the concept is not of much
help, and it is mainly of historical
value. As quoted above \cite{sorres,sorchu}, it has never
led to new answers in treating transport properties.


\begin{center}
\section*{\large{\bf Ab Initio RESULTS FOR THE WIND FORCE}}
\label{results}
\end{center}

{\it Ab initio} results for the wind valence in a number of FCC and
BCC metals are given in Table \ref{data}. The majority of the results
presented apply to self-electromigration. For impurity migration
in a metal  M the system is indicated as M(I). The calculation
for a HCP metal such as Zr was carried out by using the corresponding
FCC structure. Note, that Zr occurs in the BCC column as well. This is
done because Zr undergoes a structural phase transition from
HCP to BCC on approaching the melting temperature ${\it T}_{\rm m}$.
All $Z_{{\rm wind}}$ values are at $0.9 {\it T}_{\rm m}$.
If one is interested in the value at another temperature $T$,
one just has to look up the resistivity values at $0.9 {\it T}_{\rm m}$
and $T$, and to substitute them in the following equality
\begin{equation}
Z_{\rm wind}(T) = Z_{\rm wind}(0.9 {\it T}_{\rm m}) \rho(0.9 {\it T}_{\rm m})/\rho(T).
\end{equation}

The agreement with the limited experimental values \cite{penney64,EFW96} 
appears to be quite reasonable, particularly considering that no
self-consistent potentials were available. Further we want to draw the attention to
three typical features. First one sees that both positive and negative
wind valences are found. This is a result of the detailed
electronic structure of the system, and it could never come from a
jellium calculation, leading always to a negative wind valence.
Secondly, it is intersting to
observe, that two impurities in the same metal, {\it e.g.} Zr and Tc in
Nb, have different signs of the wind valence. At first sight
this is counterintuitive, if one considers the current flow in
the host metal as being decisive for the sign of the wind valence.
However, in the formalism all scattering effects are taken into account.
A typical multiple-scattering effect of an impurity in a metal
is the back-scattering, which can be much different for different
impurities.
Finally, $Z_{{\rm wind}}$ appears to be rather
small for quite a number of metals. This means, that a measurement
of the effective valence $Z^{*}$ will be valuable in the
determination of the direct valence.


\begin{center}
\section*{\large{\bf CONCLUSIONS AND PERSPECTIVES}}
\label{conclusions}
\end{center}

In discussing the Hellmann-Feynman theorem in relation to the driving force in
electromigration, it is made clear that the latter requires the tools
developed in statistical mechanics, to which
the theorem does not give a clue. The theorem can be considered as
being only of historical value, in providing confidence in the
force operator as a proper quantum mechanical entity.

A reliable expression for the wind force is available, being applicable 
in real systems, such as metals and dilute alloys. Strong evidence is available
for cancellation of the direct force on an impurity embedded in an electron
gas, so that only the wind force is left, although some
results do not confirm this yet. The influence
of electron-phonon interaction on this cancellation is still not entirely
clear.

Local concepts are illustrated to cover real entities, but
in the actual theory applied in practical calculations they do not enter. That is why
these concepts never have led to new answers in dealing with
transport properties. Reality is completely covered by a
standard calculation accounting for all scattering effects.
The most powerful method available at present is the KKR-Green function theory.
The {\it ab initio} obtained wind forces in a number of FCC and BCC
metals can be of practical value. They can be used
also in deciding about the direct force, because in several
systems the wind force turns out to be very small.

Up to now results are available for bulk electromigration only. For
substitutional electromigration this implies a configuration of a migrating
atom exchanging position with a neighboring vacancy. We want to
stipulate that the formalism allows for treatment of much more
flexible configurations, which could simulate both surface electromigration
and migration along a grain boundary. The complete electromigration
defect accounted for consists of 20 and 16 scattering centers in the
FCC and BCC structures resp. This defect consists of the migrating atom,
including its initial and final position, and all neighboring
perturbed metallic atoms around these two positions. By omitting 
more and more atoms in the defect one can gradually generate
a void. It is well possible to calculate the wind force on an atom
moving in such a void along the inner metallic surface. Such
a configuration would simulate surface electromigration. Other
configurations can be designed to simulate a grain boundary. It is
most promising to look to such configurations in the
near future.
\\
\section*{\large{\bf ACKNOWLEDGEMENT}}
This work was sponsored by the
National Computing Facilities Foundation (NCF) for
the use of supercomputer facilities, with financial support from the
Nederlandse Organisatie voor Wetenschappelijk Onderzoek (Netherlands
Organization for Scientific Research, NWO).
\\
\def\refname{\large{\bf REFERENCES}}

\end{document}